\documentclass[prl,reprint,amsmath,amssymb,notitlepage]{revtex4-1}

\usepackage[utf8]{inputenc}
\usepackage{hyperref,braket,graphicx,color,physics,mathtools,amsfonts}

\begin{document}

\textbf{Comment on ``Strong Measurements Give a Better Direct Measurement of the Quantum Wave Function''}
\\

In a recent theoretical result~\cite{vallone2016strong}, Vallone and Dequal (VD) claim that a better direct measurement of a quantum wave-function can be obtained using strong measurements compared to the original weak measurement proposal~\cite{lundeen2011direct}. In this Comment, we show that certain entities in Eq.~(S6) of their supplemental materials on which the central claim of their result is based have no operational existence and are therefore not experimentally measurable. 

The real and imaginary parts of the unknown wave-function to be measured are expressed as a function of
``probabilities'' in Eq.~(S6) of the supplemental materials of the aforesaid Letter. This equation is crucial
to the further development of the Letter (through Equations (4) and (5) in the main text) and is intended to be the ``strong'' analogue of Equations (4), (6)
and (7) in Ref.~\cite{lundeen2011direct}. However, the entities $ P_+^{(x)} $, $ P_-^{(x)} $, $ P_1^{(x)} $,
$ P_L^{(x)} $, and $ P_R^{(x)} $ on the right hand side of Eq.~(S6) are calculated using the
\textit{unnormalized} pointer wave-function $ \ket{\varphi}_\mathcal{P} $ (Eq.~(S2)). Consequently, they end
up being the joint probabilities:
\begin{equation*}
P_+^{(x)} = \braket{\varphi}{+} \braket{+}{\varphi}_\mathcal{P}, \quad \text{and} \quad P_-^{(x)} = \braket{\varphi}{-} \braket{-}{\varphi}_\mathcal{P}. 
\end{equation*}  
Not surprisingly, $P_+^{(x)} + P_-^{(x)} = \braket{\varphi}{\varphi}_\mathcal{P} \neq 1 $ (identical result
holds for the other dichotomic observables). On the other hand, any experiment to measure probabilities on
a post-selected quantum ensemble renders conditional probabilities since normalization is inherently built
into the statistics of these experiments. This is witness to the fact that normalization in quantum mechanics
is not a mathematical triviality but is an operational truth~\cite{von1955mathematical}. The respective
conditional probabilities are:
\begin{equation*}
P_{\mathcal{C}+}^{(x)} = \frac{\braket{\varphi}{+} \braket{+}{\varphi}_\mathcal{P}}{\braket{\varphi}{\varphi}_\mathcal{P}}, \quad \text{and} \quad P_{\mathcal{C}-}^{(x)} = \frac{\braket{\varphi}{-} \braket{-}{\varphi}_\mathcal{P}}{\braket{\varphi}{\varphi}_\mathcal{P}}. 
\end{equation*}
Similar expressions hold for the other conditional probabilities $P_{\mathcal{C}1}^{(x)}$, $ P_{\mathcal{C}0}^{(x)} $, $ P_{\mathcal{C}L}^{(x)} $ and $ P_{\mathcal{C}R}^{(x)} $. In order to obtain joint probabilities from these conditional probabilities, the factor
$ \braket{\varphi}{\varphi}_\mathcal{P} $ needs to be multiplied~\cite{aharonov2005quantum}
to the experimentally obtained results of probability measurement. From Eq.~(S2), this factor
can be calculated to be:
\begin{equation}
\braket{\varphi}{\varphi}_\mathcal{P} = \frac{1}{d} \big[ |\tilde{\psi}|^2 - 2(1 - \cos\theta)\tilde{\psi}\Re(\psi_x) + 2(1 - \cos\theta) |\psi_x|^2  \big].
\end{equation}
The above factor contains the unknown wave-function $ \psi_x $ and is therefore not known to the experimenter and eventually cannot be multiplied. It also leads to the consequence that there is no uniform normalizing factor for the whole range of $x$ (whether discrete or continuous) over which the probabilities are calculated and thus 
the final estimated wave-function cannot be obtained from a simple form as given by Eq.~(1) in Ref.~\cite{vallone2016strong}. Also, one cannot express the real and imaginary parts of the normalized wave-function in terms of the conditional probabilities. Doing so will only give a ratio of terms containing the unknown wave-function at point $ x $ in the numerator and the denominator. However, the same can be done in the weak limit. In the weak limit, when $ \theta \approx 0 $, $ \cos(\theta) \rightarrow 1 $ and the dependence of the normalization factor on $ \psi_x $ vanishes making it uniform: $ |\tilde{\psi}|^2/d $. 

The above issue is absent in the original proposals because the sum of the weak values of projectors
belonging to the complete eigenbasis \textit{is} the complete normalized wave-function. This is due to the particular choice of post-selection on
a zero momentum eigenstate (for the 
continuous case) or on the $ \ket{+} $ state corresponding to the Hadamard transform of $ \ket{0} $
in a d-dimensional basis (discrete analogue of the continuous case). Therefore, $ \langle \Pi_x \rangle_W $
in Eq.~(4) of Ref.~\cite{lundeen2011direct} \textit{is} the continuous variable normalized wave-function
at position $ x $. Similarly, $ \ket{\psi} = \sum_{a}^{} \langle \Pi_a \rangle_W \ket{a}  $ in Eq.~(7) of 
Ref.~\cite{lundeen2011direct} \textit{is} the complete d-dimensional normalized wave-function. As is 
obvious, the weak values are obtained from the experiments.

To conclude, the direct strong measurement of the quantum wave-function, as proposed by VD, has no operational basis since it violates the basic tenets of quantum theory by expressing the unknown wave-function in terms of probabilities calculated from an unnormalized pointer state which cannot be measured experimentally. Therefore, this proposal, as such, does not do justice to the claim it posits.

We are grateful to Arun Kumar Pati for helpful discussions. This work was supported by the Department of Atomic Energy, Government of India.
\\
\\
\\
Varad R.~Pande$^{a,1}$, Som Kanjilal$^{b,2}$ and Debmalya Das$^{a,3}$\\
\\
$^{a}$Harish-Chandra Research Institute (HRI), HBNI, Chhatnag Road, Jhunsi, Allahabad, India 211019\\
\\
$^{b}$Center for Astroparticle Physics and Space Sciences, Bose Institute, Kolkata, India 695016 \\
\\
$^{1}$varadrpande@gmail.com\\
$^{2}$somkanji@jcbose.ac.in\\
$^{3}$debmalyadas@hri.res.in
 
\bibliography{DSSM_comment}
\bibliographystyle{apsrev4-1}

\end{document}